\documentclass{article}
\usepackage{sprocl1,epsfig}
\usepackage{wrapfig}

\begin{document}

%\initfloatingfigs

%\input{psfig}
%\bibliographystyle{unsrt}    % for BibTeX - sorted numerical labels by order of
                             % first citation. 

\newcommand{\lapp}{\mbox{\raisebox{-.6ex}{$\,\stackrel{\textstyle <}{\sim}\,$}}}

%\begin{document}

\title{DISCOVERY LIMITS FOR EXTRA GAUGE BOSONS IN $e^+e^- \to \nu 
\bar{\nu} \gamma$\footnote{Supported in part by the Natural Sciences 
and Engineering Research Council of Canada.}}

\author{ STEPHEN GODFREY, PAT KALYNIAK, BASIM KAMAL}

\address{Ottawa-Carleton Institute for Physics \\
Department of Physics, Carleton University, Ottawa CANADA, K1S 5B6}

\author{ARND LEIKE}

\address{LMU, Sektion Physik, Theresienstr.\ 37 , D-80333 M\"{u}nchen, Germany}

\maketitle
\abstracts{
We study the sensitivity of the process $e^+ e^- \to 
\nu\bar{\nu}\gamma$ to extra gauge bosons, particularly $W'$ bosons.  
Depending on the model, evidence for extra $W$ bosons 
in this process can be detected for $W'$ masses up to several TeV.  
}
  
\section{Introduction}

Extra charged and neutral gauge bosons are a feature of many models of 
physics beyond the standard model~\cite{c-g}.  
If discovered they would represent 
irrefutable  proof of new physics, most likely that the Standard Model 
gauge group must be extended.  Indirect 
limits exist on extra gauge bosons from precision electroweak 
measurements and direct limits from searches at high energy 
colliders, with the highest current 
limits from the Tevatron Collider at 
Fermilab~\cite{pdb}.  The Large Hadron Collider at CERN will extend the 
search for $W'$'s and $Z'$'s to several TeV~\cite{c-g}.  Precision measurements of cross sections and asymmetries for 
$f\bar{f}$ final states in high energy $e^+e^-$ collisions can reveal 
evidence for $Z'$'s ranging from several TeV to tens of TeV depending 
on the model~\cite{c-g}.  However, there are no analogous limits for $W'$'s.

Hewett suggested that the process $e^+e^- \to \nu 
\bar{\nu}\gamma$ is sensitive to extra $W$ bosons in addition to 
$Z'$'s~\cite{hewett}.  
The standard model reaction proceeds via s-channel $Z$ 
exchange and t-channel $W$ exchange.  In extended models the amplitudes 
are modified by both s-channel $Z'$ and t-channel $W'$ exchange.

In this contribution we present the expected discovery reach 
of this process for 
several models with extended gauge groups.  We also consider the 
sensitivity of our results to different luminosities and to a 
small systematic error.

We considered a number of models with $W'$'s.  
The Left-Right Model (LRM)~\cite{LRM}
is based on the gauge group $SU(2)_L 
\times SU(2)_R \times U(1)_{B-L}$ giving rise to a $W'$ and a $Z'$.  
The $W'$ is right handed and we assume massless Dirac
neutrinos. The model is parametrized by the 
ratio of the coupling constants of the two $SU(2)$ gauge groups, 
$\kappa =g_L/g_R$, which we vary over the range 
$0.55 \lapp \kappa \lapp 2.0$~\cite{hewett,kappa}.  $M_{Z'}$ and $M_{W'}$
are 
related by 
$M^2_{Z'}/M^2_{W'} = \rho \kappa^2 /(\kappa^2 - \tan^2\theta_W)$
where $\rho$ describes the Higgs content of the model. We assume $\rho =
1$, corresponding to Higgs doublets. 
The Un-Unified Model (UUM)~\cite{UUM} employs the gauge symmetry 
$SU(2)_q \times SU(2)_l \times U(1)_Y$, with left-handed quarks
and leptons transforming as doublets under their respective $SU(2)$
groups.  We parametrize the UUM by an angle $\phi$,
which represents
the mixing of the charged gauge bosons of the two $SU(2)$ groups, and by
$M_{W'}$, taken to be equal to $M_{Z'}$.
 The existing constraint on $\phi$ is $0.24 \lapp
\sin \phi \lapp 0.99$~\cite{BR}.
Kaluza-Klein Excitations (KKM) exist in models containing 
large extra dimensions~\cite{ark}.  The fermion coupling of the first KK 
excitations is enhanced by a factor of $\sqrt{2}$.
Finally we include  Sequential Standard Models (SSM) which
are not true models but have become a 
standard benchmark used to compare the discovery reach.  We consider a 
SSM $W'$ with no $Z'$ (SSM1)
 and a SSM with both a $Z'$ and $W'$, of equal
mass (SSM2).

\section{Calculation and Results}

\begin{wrapfigure}{r}{2.3in}
\centerline{\epsfig{file=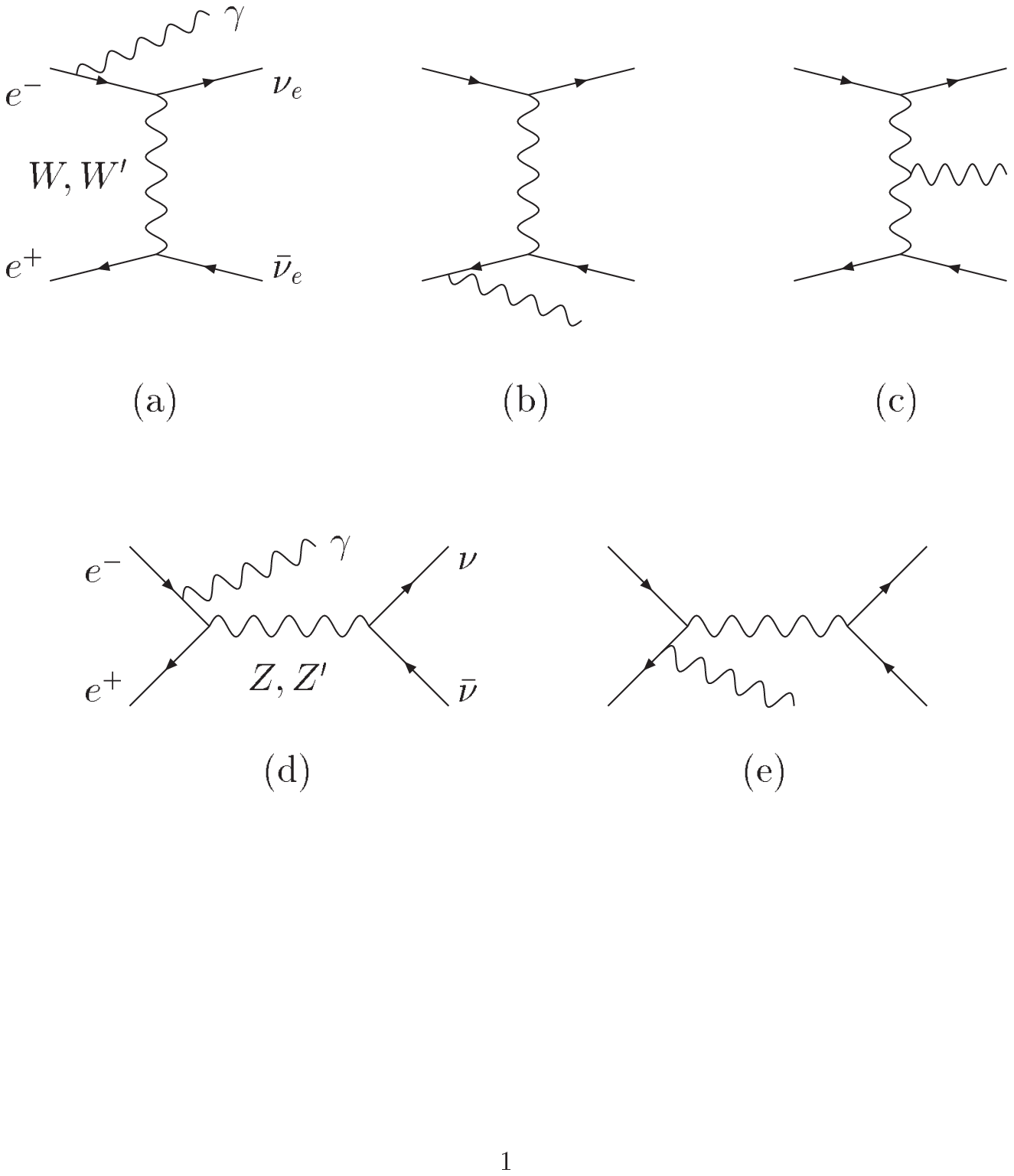,width=2.0in,clip}}
\caption{The Feynman diagrams contributing to the process $e^+e^-\to 
\nu\bar{\nu}\gamma$
}
\end{wrapfigure}
\quad The process we are studying is
\begin{equation}
e^+e^- \to \gamma \nu \bar{\nu}
\end{equation}
where the neutrinos only manifest themselves as missing energy and 
momentum.  The process is described by the Feynman diagrams of Fig. 1.

Calculation of the cross section is
relatively straightforward.  We did the calculation in a 
number of ways to give us independent checks. 
As input, we take $M_W = 80.33$ GeV, $M_Z = 91.187$ GeV, 
$\sin^2\theta_W = 0.23124$, $\alpha=1/128$, $\Gamma_Z=2.49$ GeV.

We included the following kinematic cuts to reflect finite detector 
acceptance: $E_\gamma > 10$~GeV and $10^o \leq \theta_\gamma \leq 
170^o$ where $\theta_\gamma$ is the 
angle of the photon relative to the beam.  
Our process is relatively background free with the most 
dangerous background coming from Brems\-strah\-lung events of
Bhabha-scattering with the electron and the positron lost down 
the beam pipe.  This background can be eliminated by the kinematic 
constraint
$p_T^\gamma > \sqrt{s} \sin\theta_\gamma \sin\theta_v 
/(\sin\theta_\gamma + \sin\theta_v)$  where $\theta_v$ is the minimum 
angle for veto detectors to observe activity;
we take $\theta_v =25$~mrad.

Fig. 2 shows the total unpolarized and 100\% left and right polarized 
cross sections ($\sigma_L$ and $\sigma_R$) for the SM, LRM (
$\rho=\kappa=1$), 
UUM ($\sin\phi=0.6$), SSM and KK models, all with
$M_{W'}=750$~GeV.  The peaks are due to $Z'$'s.  
At large $\sqrt{s}$ the $t$-channel dominates so, for right-handed
polarization, the LRM exhibits the largest deviation from the SM.
Polarization will be seen to be an important tool 
for discriminating between models and to constrain couplings.
The enhanced couplings of the KKM yield striking results in general.

Fig. 3 shows the differential cross section, $d\sigma/dE_\gamma$, 
for 100\% polarized electrons and $\sqrt{s}=500$~GeV.   To 
gauge the relative statistical significance of the different kinematic 
regions we plot the deviations between the SM and extended model 
differential cross sections divided by the square root of the SM 
differential cross section (which is proportional to the statistical 
error and would be normalized by the integrated luminosity).
The peak at large $E_\gamma$ is due to the radiative 
return to the $Z^0$ and is insensitive to extra gauge bosons.
To eliminate the $Z^0$ peak, which contributes nothing to the 
sensitivity to $W'$'s and $Z'$'s, we impose the additional cut 
$E_\gamma^{max} < \frac{\sqrt{s}}{2}(1-M^2_{Z^0}/s)-6\Gamma_{Z^0}$.

\begin{wrapfigure}{r}{2.2in}
\centerline{\epsfig{file=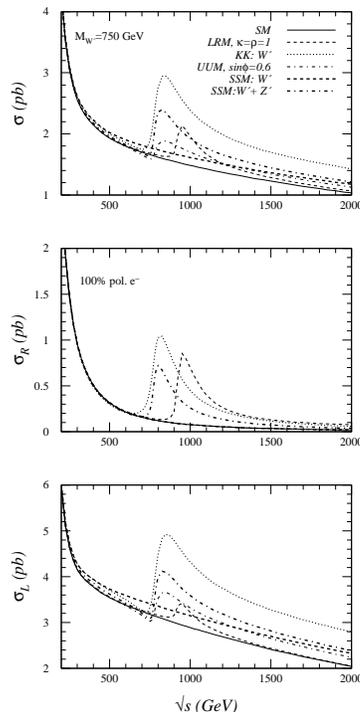,width=2.4in,clip}}
\caption{The cross section versus $\sqrt{s}$
for several models for unpolarized (top),
right-handed (middle) and left-handed (bottom) $e^-$ beam.}
\end{wrapfigure}
\quad
Fig. 3 further demonstrates the importance of polarization 
for $W'$ searches.  For example, the LR model has 
right-handed couplings and so does not deviate significantly from the SM
for left-handed polarization but does for
right-handed polarization.  In contrast, a left-handed SSM $W'$
contributes only to $\sigma_L$  while the KK
model 
contributes to both.  Note that the right-handed cross sections are 
significantly smaller in magnitude than the left-handed cross 
sections.  So although the deviations are far more pronounced for 
right-handed couplings,  they are not necessarily more statistically 
significant for polarization below 100\%.  The large difference 
between $\sigma_L$ and $\sigma_R$ also means that 
unpolarized cross sections are dominated by the left handed 
contributions.

We examined a number of observables:
total cross section, $\sigma_L$ and 
$\sigma_R$, Left-Right asymmetry ($A_{LR}$), Forward-Backward 
asymmetries, and binned photon energy and photon angular distributions.
Generally, the $d\sigma/dE_\gamma$ distributions for polarized $e^-$
were most sensitive to new gauge bosons.  However, in many cases the total 
and polarized
cross sections with the $E_\gamma$ cut and $A_{LR}$ were comparable in 
sensitivity. 
% To find the discovery
%reach values using $d\sigma/dE_\gamma$ we divided into 10 bins 
%the energy range $E_\gamma^{min}<E_\gamma < E_\gamma^{max}$ 
%where $E_\gamma^{min} = \sqrt{s} \sin\theta_v /(1 + \sin\theta_v)$ 
% as dictated by the $p_T^\gamma$ cut.
We considered two integrated luminosities to see how the limits 
varied.   We obtained limits 
by  calculating the $\chi^2$ from the difference between the extended 
model and the standard model and dividing by the statistical error 
assuming the non-standard cross section was measured.
 We found limits based on 
the statistical errors alone and then included a 2\% systematic error 
combined in quadrature with the statistical error.  Finally, we 
considered 100\% and 90\% electron polarization.
One sided 95\% C.L.  discovery limits 
for the various models, assuming 90\% polarization,
are summarized in Table 1 for $\sqrt{s}=$ 0.5, 1.0, and 1.5 TeV with 
the integrated luminosities given in the table. We present the discovery
limits for the polarized cross sections. In the cases 
that the energy distribution was most sensitive,
the discovery limit difference was about 50 GeV.

\begin{figure}[ht]
\begin{minipage}[b]{2.5in}
\centerline{\epsfig{file=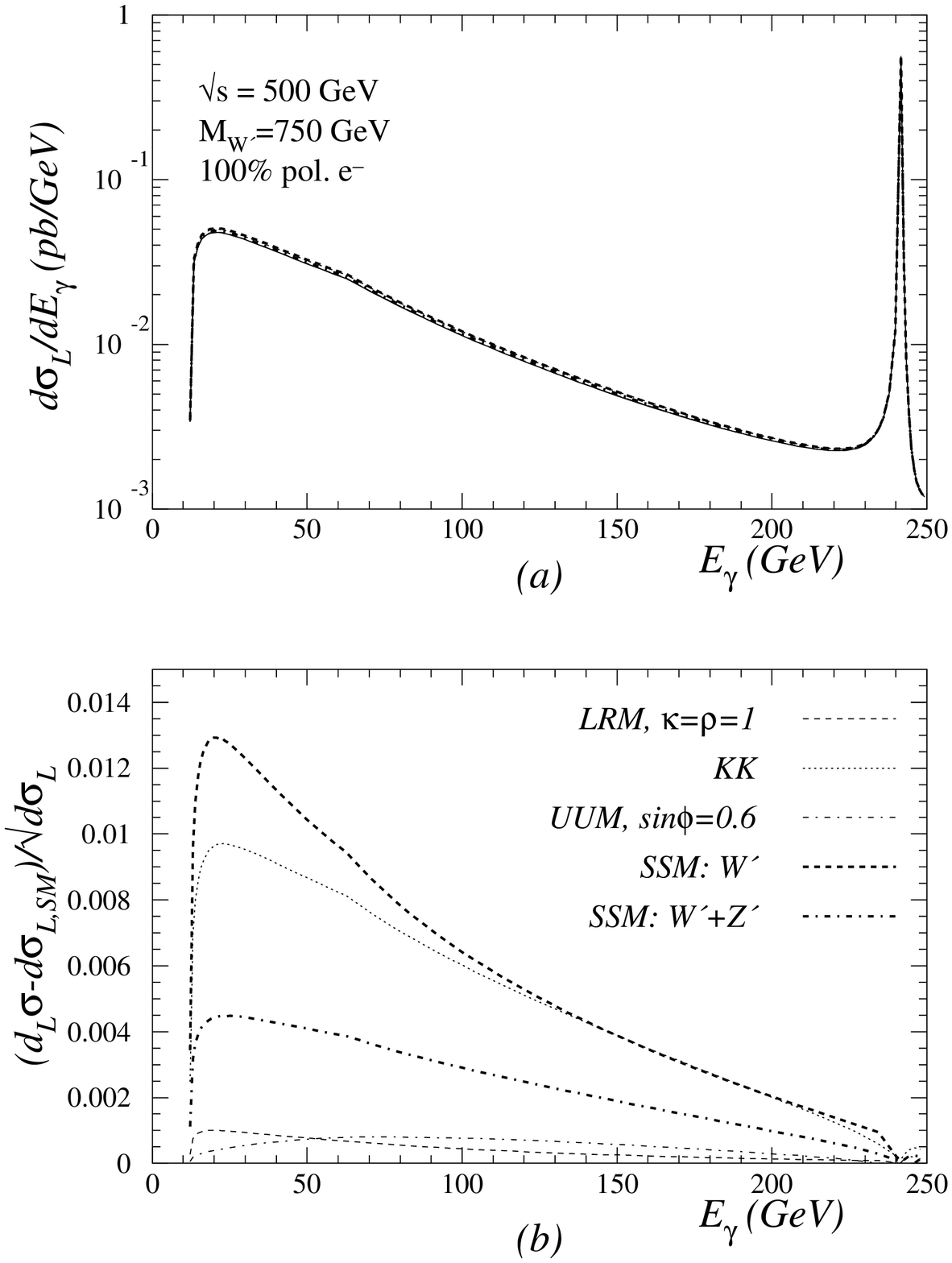,width=2.5in,clip}}
\end{minipage}
\begin{minipage}[b]{2.5in}
\centerline{\epsfig{file=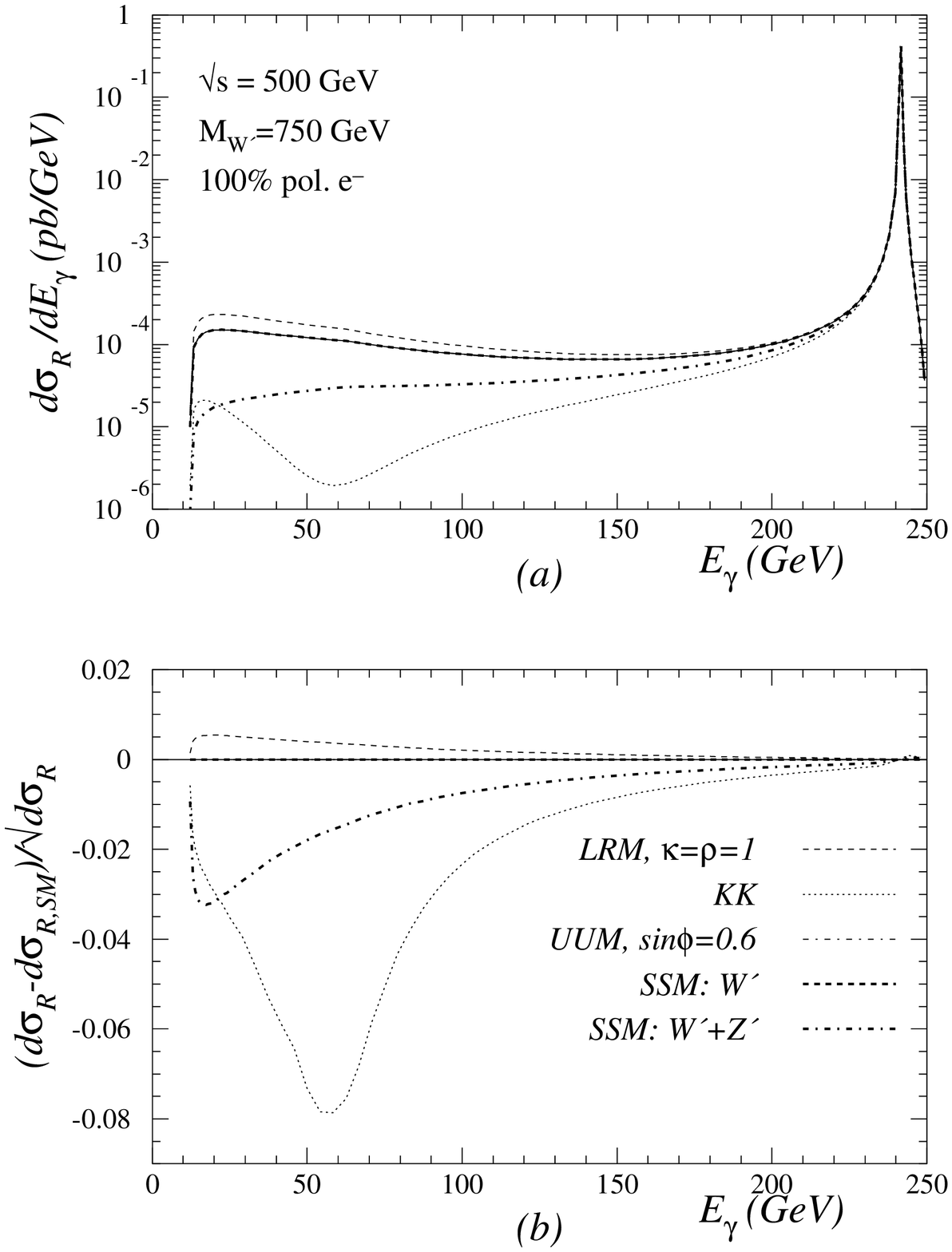,width=2.5in,clip}}
\end{minipage}
\caption{The Left and Right polarized differential cross sections (top) and 
the relative statistical significance of the
deviation from the SM (bottom) versus photon energy.}
\end{figure}

Depending on the model, the search reach for $W'$'s can be quite 
substantial, especially for the high luminosity scenario.  However, 
including even a relatively small 
systematic error of 2\%  reduces the limits from the total cross sections 
substantially. The limits obtained from the energy distribution were not 
degraded nearly as much by the systematic error. However, our results 
which include systematic errors for distributions
can only be considered approximate without a proper detector simulation. 
It is clear that systematic errors as large as 2\% are likely to dominate
as, once they 
are included, both luminosities lead to similar limits.  We also note 
that for several cases $\sigma_R$ with 
100\% polarization yields the highest limits.  However, with 90\% 
polarization $\sigma_L$ 
generally yields the highest limits.  This is a consequence of the 
larger left-handed cross section dominating the right-handed 
cross section with even a small pollution of $e^-_L$ in the 
$e^-_R$ beam.  To some extent this can be overcome by polarizing both 
the $e^-$ and $e^+$ beams, effectively increasing the net 
polarization.

\begin{table}[t]
\caption{$W'$ 95\% C.L.\ discovery 
limits obtained in the SSM1 ($W'$), SSM2 ($W'$ and $Z'$), LRM
($\kappa=\rho=1$),
UUM ($\sin\phi=0.6$), and the KKM.  We assume 90 percent $e^-$
polarization 
and use $1/2$ the stated unpolarized luminosity for the left and right cases.}
\vspace{0.4cm}
\begin{center}
\begin{tabular}{lllllll}
\hline
$\sqrt{s}$   & Model & Observable & 50 fb$^{-1}$ & 500 fb$^{-1}$ 
	& 50 fb$^{-1}$ & 500 fb$^{-1}$ \\
(GeV)	&	& 	& 	&	& + 2\% sys & + 2\% sys \\
\hline
%\vspace 0.1cm
%500 & SSM ($W'$) & $d\sigma_L/dE_\gamma$ & 2.85 & 5.05 & 1.65 & 1.72 \\
%   & SSM ($Z'$) &  $d\sigma_R/dE_\gamma$ & 2.50 & 2.75 & 1.42 & 1.86 \\
%   & LRM 	& $d\sigma_R/dE_\gamma$  & 0.92 & 1.40 & 0.86 & 1.10 \\
%   & UUM 	& $d\sigma_L/dE_\gamma$ & 0.64 & 1.85 & 0.55 & 0.56 \\
%   & KKM	& $d\sigma_L/dE_\gamma$ & 3.10 & 5.58 & 1.75 & 1.85 \\
%\hline
%\vspace 0.1cm
%1000 & 	&	&  200 fb$^{-1}$ & 500 fb$^{-1}$ 
%	& 200 fb$^{-1}$ & 500 fb$^{-1}$ \\
%	&	& 	& 	&	& + 2\% sys & + 2\% sys \\
%   & SSM ($W'$) & $d\sigma_L/dE_\gamma$ & 4.96 & 6.25 & 2.15 & 2.19 \\
%   & SSM ($Z'$) &  $d\sigma_L/dE_\gamma$ & 3.18 & 3.95 & 1.56 & 1.58 \\
%   &		&  $d\sigma_R/dE_\gamma$ & 2.47 & 3.06 & 2.02 & 2.22 \\
%   & LRM 	& $d\sigma_R/dE_\gamma$  & 1.53 & 1.78 & 1.31 & 1.38 \\
%   & UUM 	& $d\sigma_L/dE_\gamma$ & 1.24 & 1.29 & --- & --- \\
%   & KKM	& $d\sigma_L/dE_\gamma$ & 5.50 & 6.95 & 2.27 & 2.31 \\
%\hline
%1500 & 	&	&  200 fb$^{-1}$ & 500 fb$^{-1}$ 
%	& 200 fb$^{-1}$ & 500 fb$^{-1}$ \\
%	&	& 	& 	&	& + 2\% sys & + 2\% sys \\
%   & SSM ($W'$) & $d\sigma_L/dE_\gamma$ & 5.55 & 6.96 & 2.50 & 2.54 \\
%   & SSM ($Z'$) &  $d\sigma_L/dE_\gamma$ & 3.65 & 4.46 & 1.96 & 1.97 \\
%   &		&  $d\sigma_R/dE_\gamma$ & 2.76 & 3.40 & 2.32 & 2.55 \\
%   & LRM 	& $d\sigma_R/dE_\gamma$  & 1.86 & 2.14 & 1.64 & 1.77 \\
%   & UUM 	& $d\sigma_L/dE_\gamma$ & 1.76 & 1.84 & --- & --- \\
%   & KKM	& $d\sigma_L/dE_\gamma$ &  &  & 2.54 & 2.59 \\
500 & SSM1 & $\sigma_L$ & 2.4 & 4.25 & 1.0 & 1.0 \\
   & SSM2 &  $\sigma_L$ & 1.80 & 3.25 & 0.5 & 0.5 \\
   & LRM 	& $\sigma_R$  & 0.8 & 1.25 & 0.7 & 0.75 \\
   & UUM 	& $\sigma_L$ & 0.6 & 2.0 & 0.55 & 0.55 \\
   & KKM	& $\sigma_L$ & 2.60 & 4.65 & 1.0 & 1.0 \\
\hline
%\vspace 0.1cm
1000 & 	&	&  200 fb$^{-1}$ & 500 fb$^{-1}$ 
	& 200 fb$^{-1}$ & 500 fb$^{-1}$ \\
	&	& 	& 	&	& + 2\% sys & + 2\% sys \\
   & SSM1 & $\sigma_L$ & 4.15 & 5.25 & 1.25 & 1.25 \\
   & SSM2 &  $\sigma_L$ & 3.15 & 4.1 & 0.95 & 0.95 \\
   & LRM 	& $\sigma_R$  & 1.35 & 1.6 & 1.05 & 1.05 \\
   & UUM 	& $\sigma_L$ & 1.2 & 2.35 & 1.05 & 1.05 \\
   & KKM	& $\sigma_L$ & 4.55 & 5.75 & 1.0 & 1.0 \\
\hline
1500 & 	&	&  200 fb$^{-1}$ & 500 fb$^{-1}$ 
	& 200 fb$^{-1}$ & 500 fb$^{-1}$ \\
	&	& 	& 	&	& + 2\% sys & + 2\% sys \\
   & SSM1 & $\sigma_L$ & 4.65 & 5.8 & 1.45 & 1.45 \\
   & SSM2 &  $\sigma_L$ & 3.45 & 4.45 & 1.45 & 1.45 \\
   & LRM 	& $\sigma_R$  & 1.7 & 1.9 & 1.4 & 1.45 \\
   & UUM 	& $\sigma_L$ & 1.75 & 1.8 & 1.4 & 1.4 \\
   & KKM	& $\sigma_L$ & 5.05 & 6.45 & 1.45 & 1.45 \\
\hline
\end{tabular}
\end{center}
\end{table}

\section{Summary and Outlook}

In this contribution we demonstrated the usefulness of the process 
$e^+e^-\to \nu\bar{\nu} \gamma$ for $W'$ searches.  The results are 
sensitive to the models so that, if there is evidence for an extended 
gauge sector, this process could be used to help identify the model.  In 
particular, an analysis of $Z'\nu\nu$ couplings and $W'$ 
identification will be presented in a forthcoming publication.

%\section*{Acknowledgments}

\end{document}